\documentclass[11pt]{article}
\usepackage{amsmath}

\begin{document}
\topmargin -1.4cm
\oddsidemargin -0.8cm
\evensidemargin -0.8cm
\def\g#1{{\scriptstyle (\! #1 \! )}}
\def\Sg#1{{\scriptstyle [\! #1 \! ]}}
\def\Ng#1{{\scriptstyle ~\! #1 \! ~}}
\def\gg#1{{\scriptscriptstyle (\! #1 \! )}}
\def\Sgg#1{{\scriptstyle [\! {\scriptscriptstyle #1} \! ]}}
\def\Ngg#1{{\scriptstyle ~\! {\scriptscriptstyle #1} \! ~}}
\def\IUg[#1]{ \mathsf{I}^{\scriptstyle [\! #1 \! ]}{}}
\def\IDg[#1]{ \mathsf{I}_{\scriptstyle [\! #1 \! ]}{}}
\def\IUgg[#1]{\mathsf{I}^{\scriptscriptstyle [\! #1 \! ]}{}}
\def\IDgg[#1]{\mathsf{I}_{\scriptscriptstyle [\! #1 \! ]}{}}
\def\xk{{\mathbf{x},k}}
\def\xp[#1]{{\mathbf{x}\!+\!#1}}
\def\xm[#1]{{\mathbf{x}\!-\!#1}}
\def\yp[#1]{{\mathbf{y}\!+\!#1}}
\def\ym[#1]{{\mathbf{y}\!-\!#1}}
\def\DEFxmk{{\mathbf{x}-a\mathbf{e}_k,k}}
\def\DEFxpk{{\mathbf{x}+a\mathbf{e}_k,k}}
\def\Rgen[#1]#2#3{{ 
        D^{ {\scriptstyle (\! #1 \! )}#2 }_{~~~\cdot #3} }}
\def\gUP[#1]#2{{   g_{{[\! #1 \! ]}}^{{#2 }} }}
\def\gDOWN[#1]#2{{ g^{{[\! #1 \! ]}}_{{#2 }} }}
\def\ThreeJUP[#1]#2{  \left|{}_{[\!#1\!]}^{#2} \right| }
\def\ThreeJDOWN[#1]#2{\left|{}^{[\!#1\!]}_{#2} \right|}
\def\ThreeJ#1#2#3#4#5#6{{\left(\!
\begin{array}{ccc} #1 & #2 & #3 \\
                   #4 & #5 & #6
\end{array} \! 
\right)}}
\def\SixJ#1#2#3#4#5#6{{\left\{\!
\begin{array}{ccc} #1 & #2 & #3 \\
                   #4 & #5 & #6
\end{array} \! 
\right\}}}
\def\DELTA#1{{
     \!\begin{array}{c}\setlength{\unitlength}{.5 pt}
     \begin{picture}(35,25)
        \put(15, 0){\line(0,1){10}} \put(20, 0){\line(0,1){10}}
        \put(15, 0){\line(1,0){ 5}} \put(15,10){\line(1,0){ 5}}
        \put(15,25){\line(1,0){5}}  \put(15,15){$\scriptstyle {#1}$}
        \put(15,15){\oval(20,20)[l]}\put(20,15){\oval(30,20)[r]}
     \end{picture}\end{array} \!
}}
\def\THETA#1#2#3{{
     \begin{array}{c}\setlength{\unitlength}{.5 pt}
     \begin{picture}(40,40)
        \put(18,32){$\scriptstyle {#1}$}
        \put( 0,15){\line(1,0){40}} \put(18,17){$\scriptstyle {#2}$}
        \put(20,15){\oval(40,30)}   \put(18, 2){$\scriptstyle {#3}$}
        \put( 0,15){\circle*{3}}    \put(40,15){\circle*{3}}
     \end{picture}\end{array}
}}
\def\TET#1#2#3#4#5#6{{
     \begin{array}{c}\setlength{\unitlength}{.8 pt}
        \begin{picture}(50,30)
        \put( 0,15){\line(1,-1){15}} \put(0,22){${\scriptstyle {#2}}$}
        \put( 0,15){\line(1, 1){15}} \put(0, 0){${\scriptstyle {#1}}$}
        \put( 0,15){\circle*{3}}
        \put(30,15){\line(-1, 1){15}} \put(28,20){${\scriptstyle {#4}}$}
        \put(30,15){\line(-1,-1){15}} \put(28, 2){${\scriptstyle {#5}}$}
        \put(30,15){\circle*{3}}
        \put( 0,15){\line(1,0){30}} \put(12,16){${\scriptstyle {#6}}$}
        \put(15,30){\line(1,0){25}} \put(15,30){\circle*{3}}
        \put(15, 0){\line(1,0){25}} \put(15, 0){\circle*{3}}
        \put(40, 0){\line(0,1){30}} \put(42,12){${\scriptstyle {#3}}$}
    \end{picture}\end{array}
}}


\title{\Large{{\bf The basis of the physical Hilbert space of lattice 
gauge theories}}}

\vspace{2.5cm}

\author{~\\ 
{\sc G. Burgio$^{(1)}$},
{\sc R. De Pietri$^{(2)}$},
{\sc H. A. Morales-T\'ecotl$^{(3)}$\thanks{Associate member of Abdus
Salam ICTP, Trieste, Italy.}}, \\
{\sc  L. F. Urrutia $^{(4)}$}, and 
{\sc J. D. Vergara$^{(4)}$}
\\~\\ (1) Dipartimento di Fisica, Viale delle Scienze, I-43100 Parma, Italy\\
          and INFN Gruppo collegato di Parma
\\~\\ (2) Centre de Physique Th\'eorique CNRS, 
          Case 907 Campus de Luminy\\F-13288 Marseille Cedex 9, France
\\~\\ (3) Departamento de F\'\i sica, Universidad Aut\'onoma
Metropolitana Iztapalapa, \\
          A. Postal 55-534, 09340   M\'exico, D.F. 
\\~\\ (4) Departamento de F\'\i sica de Altas Energ\'\i as, Instituto
de Ciencias Nucleares, \\
Universidad Nacional Aut\'onoma de M\'exico, A. Postal 70-543, 04510
M\'exico D.F. 
}
\date{Last revision: June 28, 1999} 
\maketitle

\begin{abstract}
Non-linear Fourier analysis on compact groups
is used to construct an orthonormal basis of the physical 
(gauge invariant) Hilbert space of Hamiltonian lattice gauge theories. 
In particular, the matrix elements of the Hamiltonian operator involved are 
explicitly computed. 
Finally, some applications and possible developments of the 
formalism are discussed.
\end{abstract}

\vfill
\begin{flushleft}
CPT-99/P.3856\\
University of Parma Preprint UPRF-99-09 \\
{\it xxx-archive:} hep-lat/9906036.\\[1cm]
\end{flushleft}

\thispagestyle{empty}


\section{Introduction}

It is generally accepted that the most powerful non-perturbative
method for the study of gauge theories is the lattice approach 
\cite{Creutz:1983}. Although most work is performed using 
Lagrangian methods, the Hamiltonian point of view offers 
some advantages \cite{Kogut:1975}. For instance, concepts like
the Wilson-Polyakov confinement test \cite{Marchesini:1981},
redefinitions of the action \cite{Menotti:1981} and the vacuum
structure are more easily and clearly formulated within the
Hamiltonian, real-time, language.

In spite of the many advances along the lines of the Hamiltonian
formulation, an important problem that remains is that the resulting
basis functions, generated by the Wilson loops, are in general
nonorthogonal and overcomplete.  This overcompleteness arises from the
well known Mandelstam identities \cite{Mandelstam:1979}. This problem
has been handled in different ways. For example, in reference
\cite{Brugmann:1991} the inner product induced by the Haar
measure  (using Creutz's algorithm \cite{Creutz:1978}), has been used to find a loop
basis of linearly independent loop states. More recently, such
difficulty was tackled by exploiting the Cayley-Hamilton relations
between matrices \cite{Schutte:1997}.

Here, we propose a different approach to describe the Hilbert space of
lattice gauge theories. This is done by using {\it harmonic analysis
on a compact Lie group} $G$, as well as the formalism of intertwining
operators. It is important to emphasize that such construction can be
carried out in an abstract way using the properties of tensor
categories, namely, the tensor category of the representations of
$G$. Consequently, the formalism can be generalized to the tensor
categories of the representations of the quantum group
$SU_q(N)$. Since when $q$ is an $n^{th}$-root of unity the number of
the irreducible representation is finite, this could provide a natural
way to obtain a finite dimensional model \cite{Burgio:1999}.

Very similar techniques have been used in the path integral approach
\cite{Aroca:1996a}, and to construct a q-deformed lattice gauge theory
\cite{Boulatov:1997}.  Our work generalizes the results obtained in
\cite{Aroca:1998} for the case of $SU(2)$ lattice gauge theory in
$2+1$ dimensions, to arbitrary dimensions and compact gauge
groups. Moreover, our procedure allows the explicit calculation of the
matrix elements of the Hamiltonian.

One of the technical difficulties in manipulating the derived formulas
is the proliferation of indexes, which turns out in quite cumbersome
expressions. The best way to gain control of the formalism is to take
advantage of the graphical methods available for performing recoupling
calculus.  The usefulness of graphical methods \cite{Yutsis:1962} in
computations involving complicated (representation theory) tensor
expressions has been particularly stressed by Citanovi\`c
\cite{Citanovic:1984} (who was the first one that emphasized the non
relevance of the particular basis, as well as the particular
normalization chosen in the definition of Clebsh-Gordan coefficients
and Wigner 3nJ-Symbol) and by Yutsis-Levinson-Vanagas
\cite{Yutsis:1962}.  These techniques have extensively been used in
the context of loop quantum gravity (see for example
\cite{DePietri:1996,DePietri:1997,DePietri:1999a} and reference
therein).

The plan of the work is the following. 
Section 2 summarizes the specific theory we will be dealing with to
calculate the matrix elements, although our results on the basis
are completely general.
Section 3 gives the basis for the gauge invariant states of the theory
and also reviews the group theoretical methods needed to achieve such
results.
In section 4 we determine the matrix elements of the Hamiltonian
operator in terms of contraction of intertwiners for arbitrary
dimension and arbitrary gauge group.  Moreover, algebraic close form
are explicitly derived for the $SU(2)$ case in 2+1 and 3+1 dimensions.
Section 5 discusses possible applications and developments, as well as
some conclusions.


\section{Hamiltonian lattice gauge theories}

The Hamiltonian formalism for lattice gauge theories in $(d+1)$
dimensions was introduced in \cite{Kogut:1975}\footnote{We use the
conventions of \cite{Marchesini:1981}.}.  To fix the notations, we are
considering a lattice gauge theory defined on a $Z^d$ periodic cubic
$d$ dimensional lattice with a continuum time.  To make contact with
the continuum theory, the lattice can be visualized as a finite subset
of points of a $d$-dimensional torus $T^d$ of principal radius $a L$
($a \in \mathbf{R}^+$), $Z^{d}=\{ \mathbf{x}\in T^d |
\mathbf{x}= a (n_1 \mathbf{e}_1+\ldots+ n_d \mathbf{e}_d) \}$, 
defined by the versors $(\mathbf{e}_1,\ldots,\mathbf{e}_d)$ and
$(n_1,\ldots n_d )\in\mathbf{Z}_L^d$. This implies that we have
$N_v=L^d$ vertices, and $N_{lk}=d\cdot L^d$ links and
$N_{P}=\frac{d(d-1)}{2}\cdot L^d$ plaquettes.

For each lattice link $(\mathbf{x},\mathbf{x}+a \mathbf{e}_k)$, one
has a gauge field variable\footnote{The link with the gauge fields
variable $A^\alpha_i(x)$ is as follow. Consider the segment
$x^i(\lambda)=(1-\lambda) \mathbf{x}^i +
\lambda (\mathbf{x}^i+a \mathbf{e}_k^i)$,
than $U_k(\mathbf{x})={\cal P}\exp \left[\int_0^1 d\lambda A_i^\alpha
T^\alpha \frac{dx^i(\lambda)}{d\lambda}
\right]$ where ${\cal P}\exp$ is the path order exponential.}
$U_k(\mathbf{x})\in G$, where $G$ is a (compact) gauge group (e.g.,
$SU(N)$).  The variables conjugated to the link variables are the
outgoing (ingoing) electric fields $E^\alpha_{\pm\! k}(\mathbf{x})$.
More precisely, $E^\alpha_{\pm\! k}(\mathbf{x})$ denotes the electric
fields outgoing (ingoing) from the lattice point $\mathbf{x}$ in the
directions $ \mathbf{e}_{\pm\!k}$. They fulfill the commutation
relations
\begin{eqnarray}
&& [E^\alpha_{+k}(\mathbf{x}),  U_j(\mathbf{y}) ]
   = \delta_{k,j} \delta^3_{\mathbf{x},\mathbf{y}} 
     U_j(\mathbf{y}) T^\alpha ~~,
\\
&& [E^\alpha_{-k}(\mathbf{x}+a \mathbf{e}_k),  U_j(\mathbf{y}) ]
   = - \delta_{k,j} \delta^3_{\mathbf{x},\mathbf{y}} 
     T^\alpha U_j(\mathbf{y}) 
~~,
\end{eqnarray}
where $T^\alpha$ are the hermitian generators of $G$ in the 
fundamental representation. 

The generators of the local gauge transformations are:
\begin{equation}
\mathcal{G}^\alpha(\mathbf{x}) 
   = D_k E^\alpha_{k}(\mathbf{x}) 
   = \sum_{k=1}^d \left(  
             E^\alpha_{+k}(\mathbf{x}) + E^\alpha_{-\!k}(\mathbf{x})
     \right)
~~.
\end{equation}
Notice that each particular gauge transformation acts on 
$U_k(\mathbf{x})\in G$
according to ($\gamma(\mathbf{x})\in G$)
\begin{equation}
  U_k(\mathbf{x}) \longrightarrow 
  U_k^\gamma(\mathbf{x}) = \gamma^{-1}(\mathbf{x}+a\mathbf{e}_k)
                           U_k(\mathbf{x}) \gamma(\mathbf{x})
~~.
\end{equation}
In discussing Hamiltonian lattice gauge theories, it is important to
consider two Hilbert spaces. The first one is the Hilbert space
$\mathcal{H}_{aux}$ composed by the square integrable functions
$\psi(U)=\psi(\{ U_k(\mathbf{x}) \})$, with respect to the unique
normalized Haar measure on the group $G$
\begin{eqnarray}
&& \int [dU] ~ |\psi(U)|^2 < \infty ~~,
   \qquad\qquad  [dU] = \prod_{\mathbf{x},k} dU_k(\mathbf{x}) ~~,
\end{eqnarray}
while the second is the physical Hilbert space of the theory 
$\mathcal{H} \subset \mathcal{H}_{aux}$ composed by the gauge invariant
square integrable functions
\begin{eqnarray} \label{eq:GAUGEINV}
&& \psi(U) = \psi(U^\gamma) =  \psi(\{ U_k^\gamma(\mathbf{x}) \})~~.
\end{eqnarray}
The Hamiltonian (defined on the two Hilbert spaces
$\mathcal{H}$ and $\mathcal{H}_{aux}$) is
\begin{equation} \label{def:HAM}
\hat{H} = \frac{1}{a} \left\{
      \frac{g^2}{2} \sum_{\mathbf{x},k} ~\sum_{\alpha\beta} 
        q_{\alpha\beta} E^\alpha_k(\mathbf{x})E^\beta_k(\mathbf{x})
     + \sum_P V(U_P) \right\}
      = H_E + H_B
\end{equation}
where $q_{\alpha\beta}$ is the Cartan metric, the sum over $P$ in 
equation (\ref{def:HAM}) ranges over all unoriented
plaquettes in the lattice space, and
\begin{equation}
  V(U_P) = \frac{1}{g^2} \left[1 - 
           \frac{U_P + U_P^*}{2\mathrm{dim}(U)} \right] 
\end{equation}
where $U_P$ is the usual plaquette variable 
defined by\footnote{We introduce the notations
${\mathbf{x}\!\pm\!k} = {\mathbf{x}\pm a \mathbf{e}_k}$. 
We will also use  ${\mathbf{x}\!\pm\!k\pm\!l} = 
{\mathbf{x}\pm a \mathbf{e}_k \pm a \mathbf{e}_l}$.}
\begin{eqnarray}
U_P &=&U_{\mathbf{x},k,l} 
      ={\rm Tr}[ U^{-1}_l(\mathbf{x}) U^{-1}_k(\mathbf{x}+a \mathbf{e}_l)
                 U_l(\mathbf{x}+a \mathbf{e}_k) U_k(\mathbf{x}) ]
\\
    &=&{\rm Tr}[ U^{-1}_l(\mathbf{x}) U^{-1}_k(\xp[l])
             U_l(\xp[k]) U_k(\mathbf{x}) ]
~~.
\nonumber
\end{eqnarray}
In this work, we will limit the analysis to the standard form of the
magnetic term derived from the Wilson action \cite{Wilson:1974}.
Such choice is not unique.
The only condition on the magnetic term potential $V(U_P)$ 
is that it has to fulfill the requirement 
$V(U_P) \simeq \frac{a^4}{2} {\rm Tr}[F_P^2]$. For example,
a very interesting alternative definition is the use of the so called
Heat-Kernel potential \cite{Menotti:1981}. The most general
one plaquette potential $V(U)$ can be written as 
\begin{equation}
   V(U)=\sum_f c_f (\chi_f[U]+ \chi_f[U]^*)~~,
\end{equation}
where the sum spans over all the irreducible representations of the group
and $\chi_f$ are the associated character functions\footnote{
The character of the unitary representation $\Rgen[f]{\alpha}{\beta}(U)$ 
of a group is its trace $\chi_f[U]=\Rgen[f]{\alpha}{\alpha}(U)$.}.
To deal with the general case one has to determine the matrix elements
of $\chi_f[U]$. In our formalism such generalization is straightforward,
it simply amounts to replace the $1$ (that denotes the fundamental 
representation) in equation (\ref{eq:completa}) with $f$. 


\section{The gauge invariant basis of the physical Hilbert space} 


\subsection{Fourier analysis on compact groups: Peter-Weyl theorem} 

A classical result of group theory is that the set 
$\mathcal{RG} = \{  \mathcal{R}^{j} ~| j\in J[G]  \}$ of all 
the matrix elements of irreducible inequivalent unitary 
representations\footnote{
We denote by $\mathcal{H}^j$ the Hilbert space in which the
irreducible representation is realized in terms of unitary operator
$T^{\g{j}}(U)$ and we suppose that a {\it preferred} orthonormal basis
has been chosen.  We will denoted by $\Rgen[j]{\alpha}{\beta}(U),$
($\alpha,\beta=1,\ldots,\mathrm{dim}(\mathcal{H}^j)$) the matrix
elements of $T^{\g{j}}(U)$ in this {\it preferred} basis.
Moreover we will use $\overline{\mathcal{R}^{j}}$ to denote
the adjoint representation, defined on the same Hilbert space
$\mathcal{H}^j$, of matrix elements $\Rgen[{}^*\!\!j]{\alpha}{\beta}(U)
=\Rgen[j]{\alpha}{\beta}(U^{-1})=\overline{\Rgen[j]{\beta}{\alpha}(U)}$
} 
$j$ of the group $G$ is a complete orthogonal basis
on the Hilbert space $\mathcal{L}^2[G,dU]$. This result,
known as the Peter-Weyl theorem \cite{Vilenkin:1993a} implies that 
any $f(U) \in \mathcal{L}^2[G,dU]$ can be expanded as
\begin{equation} \label{eq:fou}
  f(U) =\sum_{j\in J[G]} \sum_{\alpha\beta}  
        c^{\g{j}\beta}_{\alpha\cdot} ~~ \Rgen[j]{\alpha}{\beta}(U)~~. 
\end{equation}
where there is the following orthogonality property between the 
matrix functions of the irreducible representation of the group:
\begin{equation} \label{p:INT1}
  \int dU ~ \overline{\Rgen[j]{\alpha}{\beta}(U)} \Rgen[j]{\alpha'}{\beta'}(U) 
= \int dU ~ {\Rgen[j]{\beta}{\alpha}(U^{-\!1})} \Rgen[j]{\alpha'}{\beta'}(U) 
= \frac{1}{\mathrm{dim}(\mathcal{H}^j)}
  \delta_\alpha^{\alpha'} \delta^\beta_{\beta'}~~,
\end{equation}
where $\mathcal{H}^j$ is the Hilbert space on which the representation
is defined.  The general rule for performing integration over the
group yields zero unless the integrand transforms as the trivial
representation.  Clearly this construction straightforwardly
generalizes to functions on the Cartesian product of $N_{lk}$ copies
of the gauge group. In this way, the most general vector of
$\mathcal{H}_{aux}$ can be written as:
\def\SHORTCUTforC{
{c^{\g{j_\gg{1}\cdots j_\gg{N_{lk}}}
    \beta_\gg{1}\cdots\beta_\gg{N_{lk}}
}_{~~~~~~~~~~~\alpha_\gg{1}\cdots\alpha_\gg{N_{lk}}} }}
\def\SHORTCUTforCbar{
{c^{\g{j_\gg{1}\cdots j_\gg{N_{lk}}}
    \bar{\beta}_\gg{1}\cdots\bar{\beta}_\gg{N_{lk}}
}_{~~~~~~~~~~~\bar{\alpha}_\gg{1}\cdots\bar{\alpha}_\gg{N_{lk}}} }}
\begin{equation} \label{eq:vHaux}
  \psi(U) = \prod_{\mathbf{x}} \prod_{k=1}^{d} 
     \left[ \sum_{j_\mathbf{x}^k \in J[G]} 
            \sum_{\alpha_\mathbf{x}^k,\beta_\mathbf{x}^k=1}^{
                         {\mathrm{dim}(j_\mathbf{x}^k)}} 
     ~\Rgen[j_\mathbf{x}^k]{\alpha_\mathbf{x}^k}{\beta_\mathbf{x}^k}(U) 
     \right] 
    \times 
    \SHORTCUTforC
\end{equation}
At this point the Peter-Weyl theorem has given us a complete
characterization of $\mathcal{H}_{aux}$.  Notice that we are not
interested in $\mathcal{H}_{aux}$, but in its gauge invariant
subspace.  The implementation of the gauge invariance
(\ref{eq:GAUGEINV}) turns out to be ``simply'' a restriction of the
possible forms of $c$'s.  In the next subsection we will see that the
condition of gauge invariance imposes the condition that the $c$'s must
be group invariant tensors.


\subsection{Invariant tensors: the basis of the gauge invariant Hilbert space} 

The concept of invariant tensor is better expressed by the notions of 
intertwining operator\footnote{
The generalized Clebsh-Gordan coefficients of Yutsis-Levinson-Vanagas
\cite{Yutsis:1962} are just the matrix elements of these operators on
the privileged basis (introduced in the Hilbert space of the
representations).
} 
By definition, an operator ${\mathsf I}$ between the Hilbert space of
two representations, $\mathcal{R}$ and $\mathcal{R}'$ of $G$, is an
intertwining operator if $\mathsf{I}$ is a bounded operator from
$\mathcal{H}$ to $\mathcal{H}'$ such that $\mathsf{I} \cdot T(U) =
T'(U) \cdot \mathsf{I}$, $\forall U \in G$.  Now, the set of all the
intertwining operators $\mathcal{I}(\mathcal{R},\mathcal{R}')$ is a
vector subspace of the space of bounded linear operators. 

Moreover, we have the following properties between the spaces of
intertwining operators:
$\mathcal{I}(\mathcal{R}\otimes\mathcal{R}'',\mathcal{R}')
=\mathcal{I}(\mathcal{R},\mathcal{R}'\otimes\overline{\mathcal{R}''})$
and $\mathcal{I}(\mathcal{R},\mathcal{R}')$ and
$\mathcal{I}(\mathcal{R}',\mathcal{R})$ are anti-isomorphic vector
spaces. Using this duality it is natural to use the trace
function\footnote{
We are assuming that this space is finite dimensional. This is exactly
the case when the two representations $\mathcal{R}^{j}$ and
$\mathcal{R}^{j'}$ are finite dimensional.
} 
to induce an Hilbert space structure on
$\mathcal{I}(\mathcal{R},\mathcal{R}')$. In fact,
$\mathrm{Tr}[\mathsf{I}_1 \mathsf{I}_2]$ makes perfect sense when
$\mathsf{I}_1 \in \mathcal{I}(\mathcal{R},\mathcal{R}')$ and 
$\mathsf{I}_2 \in \mathcal{I}(\mathcal{R}',\mathcal{R})$.  
In terms of intertwiner operators we have the following integration 
formula for the direct product of $K$ representations:
\begin{equation}
 \int dU~ \prod_{k=1}^K 
          \Rgen[j_k]{\alpha_k}{\beta_k}(U)
= \sum_{\pi}
  \frac{\IUg[\pi]{}^{(j_1\ldots j_K)}_{\beta_1\ldots\beta_K}
        \IDg[\pi]{}_{(j_1\ldots j_K)}^{\alpha_1\ldots\alpha_K}
      }{\IUg[\pi]{}^{(j_1\ldots j_K)}_{\gamma_1\ldots\gamma_K}
        \IDg[\pi]{}_{(j_1\ldots j_K)}^{\gamma_1\ldots\gamma_K}
       }
\end{equation}
where $\IDg[\pi]{}_{(j_1\ldots j_K)}\in \mathcal{I}(\mathcal{R}^{j_1}
\otimes\ldots\otimes\mathcal{R}^{j_K},\emptyset)$,
and  $ \IUg[\pi]{}^{(j_1\ldots j_K)} \in \mathcal{I}(\emptyset,
\mathcal{R}^{j_1}\otimes\ldots\otimes\mathcal{R}^{j_K})$ is 
its adjoint intertwiner.

Having fixed the notations we are now ready to impose the condition of gauge 
invariance (\ref{eq:GAUGEINV}) to a generic vector (\ref{eq:vHaux}) of
$\mathcal{H}_{aux}$:
\begin{eqnarray} 
\psi(U) &=& \psi(U^\gamma) 
     = \psi(\{ U_k^\gamma(\mathbf{x}) \})
     = \psi(\{ \gamma^{-\!1}(\mathbf{x}+a\mathbf{e}_k)
                U_k(\mathbf{x}) \gamma(\mathbf{x}) \}) 
\label{eq:CONDc} \\ 
    &=&  \prod_{\mathbf{x}} \prod_{k=1}^{d} 
     \left[ \sum_{j_\mathbf{x}^k \in J[G]} 
            \sum_{\alpha_\mathbf{x}^k,\beta_\mathbf{x}^k=1}^{
                         {\mathrm{dim}(j_\mathbf{x}^k)}} 
     ~~ \Rgen[j_\mathbf{x}^k]{\alpha_\mathbf{x}^k}{\beta_\mathbf{x}^k}
            (\gamma^{-\!1}(\xp[k])U_k(\mathbf{x}) \gamma(\mathbf{x})) 
            \right] 
    \times \SHORTCUTforC
\nonumber \\ 
    &=& \prod_{\mathbf{x}} \prod_{k=1}^{d} 
     \Bigg[ \sum_{j_\mathbf{x}^k \in J[G]} 
            \sum_{\alpha_\mathbf{x}^k,\beta_\mathbf{x}^k=1}^{
                         {\mathrm{dim}(j_\mathbf{x}^k)}} 
       ~\Rgen[j_\mathbf{x}^k ]{\bar{\alpha}_\mathbf{x}^k}{
                               \bar{\beta}_\mathbf{x}^k}
                        (U_k(\mathbf{x})) 
       \cdot
       ~\Rgen[j_\mathbf{x}^k]{\alpha_\mathbf{x}^k }{\bar{\alpha}_\mathbf{x}^k}
              (\gamma^{-\!1}(\xp[k])) 
       ~\Rgen[j_\mathbf{x}^k]{\bar{\beta}_\mathbf{x}^k}{\beta_\mathbf{x}^k}
             (\gamma(\mathbf{x})) 
    \Bigg] \times
\nonumber \\[1mm] 
   & & \qquad\qquad
    \times \SHORTCUTforC
\nonumber
\end{eqnarray}
This gives conditions on the possible form of the
{\it generalized Fourier transform} coefficients \break
($\SHORTCUTforC$). It
implies that the coefficients of the {\it generalized Fourier
transform} of a gauge invariant function are invariant tensors under
the transformation of the gauge group associated to the vertex of the
lattice. To see this we can rewrite equation (\ref{eq:CONDc}) 
by collecting all the terms depending on the gauge transformation at 
the vertex $\mathbf{x}$. The condition of gauge invariance implies that 
at each vertex $\mathbf{x}$, the following equation must be satisfied
\begin{eqnarray} \nonumber 
\SHORTCUTforCbar &=& 
  \prod_{k=1}^d 
            \sum_{\alpha_\xm[k]^k}^{
                 {\mathrm{dim}(j_\xm[k]^k)}} 
            \sum_{\beta_\mathbf{x}^k=1}^{
                 {\mathrm{dim}(j_\mathbf{x}^k)}}  
\Bigg[ ~\Rgen[j_{\xm[k]}^k]{\alpha_{\xm[k]}^k}{~~\overline{\alpha}_{\xm[k]}^k}
(\gamma^{-\!1}(\mathbf{x}))
~\Rgen[j_\mathbf{x}^k]{\overline{\beta}_\mathbf{x}^k}{\beta_\mathbf{x}^k}
(\gamma(\mathbf{x})) \Bigg] \times \SHORTCUTforC 
\end{eqnarray}
The previous expression involves a large number of indices (like the
ones involved in the previous decomposition of our Hilbert
space). They are very cumbersome to write down, even though the
concept they express is simple.  For example, previous equation
expresses that in the case of gauge invariant
Hilbert space the $c$'s should be proportional to the generalized
Clebsh-Gordan coefficients \cite{Yutsis:1962} of all the
representations associated to the links connected to the 
the lattice site $\mathbf{x}$. That means, 
that they must be proportional to the matrix elements of 
an intertwining operator. Since this apply all the lattice site 
we have that   
\begin{equation}
\SHORTCUTforC  = 
\left[\prod_{\mathbf{x}}\sum_{\Sg{u_\mathbf{x}}} \right]
    ~
    c_{\Sg{u_{\mathbf{x}_1}},....,\Sg{u_{\mathbf{x}_{N_v}}}}
    ~
\left[
    \prod_{\mathbf{x}}     \mathsf{I}_{\mathbf{x}}^{\Sg{\pi_\mathbf{x}}}
    {}^\g{j_{\xm[d]}^1,\ldots,j_{\xm[d]}^d}_{
        \alpha_{\xm[1]}^1,\ldots,\alpha_{\xm[d]}^d}
    {}_\g{j_\mathbf{x}^1,\ldots,j_\mathbf{x}^d}^{
        \alpha_\mathbf{x}^1,\ldots,\alpha_\mathbf{x}^d}
\right]
\end{equation}
where with $\mathsf{I}_{\mathbf{x}}^{\Sg{\pi_\mathbf{x}}}$ 
we have denoted the (a choice of a possible) basis vectors,
labeled by the index $\pi_\mathbf{x}$, of the space of the 
intertwiner operators 
$\mathcal{I}( \otimes_{k=1}^d ~  \mathcal{R}^{j_{\xm[k]}^k}
,\otimes_{k=1}^d ~  \mathcal{R}^{j_\mathbf{x}^k})$ at 
each lattice site $\mathbf{x}$.
More specifically we will use the notation 
\begin{equation}
    \mathsf{I}_{\mathbf{x}}^{\Sg{\pi_\mathbf{x}}}
    {}^\g{j_{\xm[d]}^1,\ldots,j_{\xm[d]}^d}_{
        \alpha_{\xm[1]}^1,\ldots,\alpha_{\xm[d]}^d}
    {}_\g{j_\mathbf{x}^1,\ldots,j_\mathbf{x}^d}^{
        \alpha_\mathbf{x}^1,\ldots,\alpha_\mathbf{x}^d}
    =
    \langle j_{\mathbf{x}}^1 , \alpha_{\mathbf{x}}^1 |
    \otimes \ldots \otimes
    \langle j_{\mathbf{x}}^d, \alpha_{\mathbf{x}}^d  | 
    ~
    \mathsf{I}_{\mathbf{x}}^{\Sg{\pi_\mathbf{x}}}
    |j_{\xm[1]}^1 , \alpha_{\xm[1]}^1 \rangle
    \otimes \ldots \otimes
    |j_{\xm[d]}^d , \alpha_{\xm[d]}^d \rangle
\end{equation}
for their explicit matrix elements and we will denote with 
\begin{equation}
\mathsf{I}^{\mathbf{x}}_{\Sg{{\mathbf{\pi}_\mathbf{x}}}}
    {}_\g{j_{\xm[d]}^1,\ldots,j_{\xm[d]}^d}^{
        \alpha_{\xm[1]}^1,\ldots,\alpha_{\xm[d]}^d}
    {}^\g{j_\mathbf{x}^1,\ldots,j_\mathbf{x}^d}_{
        \beta_\mathbf{x}^1,\ldots,\beta_\mathbf{x}^d}
= \overline{\mathsf{I}_{\mathbf{x}}^{\Sg{\mathbf{\pi}_\mathbf{x}}}
    {}^\g{j_{\xm[d]}^1,\ldots,j_{\xm[d]}^d}_{
        \alpha_{\xm[1]}^1,\ldots,\alpha_{\xm[d]}^d}
    {}_\g{j_\mathbf{x}^1,\ldots,j_\mathbf{x}^d}^{
        \beta_\mathbf{x}^1,\ldots,\beta_\mathbf{x}^d}}
\end{equation}
the complex conjugate (adjoint) intertwining operator.

This result implies that the $c$'s coefficients factorize in a product 
of $c$'s, one for each vertex of the lattice. 
Summarizing, the application of the Peter and Weyl theorem
and the imposition of gauge invariance gives the following
description of the Hilbert space $\mathcal{H}$ in terms
of the orthogonal, so called {\it spin-network}, basis:
\begin{equation} \label{eq:SpinNetBasis}
\psi(U)= \sum_{ \vec{\jmath}, \vec{\mathbf{\pi}} }
           c_{ \vec{\jmath}, \vec{\mathbf{\pi}} }
          \psi_{ \vec{\jmath}, \vec{\mathbf{\pi}} }(U) 
       = \prod_{\mathbf{x}} \prod_{k=1}^{d} 
         \left[ \sum_{j_\mathbf{x}^k \in J[G]}
                \sum_{\mathbf{\pi}_\mathbf{x}}
         \right] \psi_{ \vec{\jmath}, \vec{\mathbf{\pi}} }(U) 
\end{equation}
where the sum over the $\mathbf{\pi}_\mathbf{x}$ ranges over a complete
labeling of the basis of the intertwiners
$\mathcal{I}(\otimes_{k}^d\mathcal{R}^{j_{\xm[k]}^k},
\otimes_{k}^d \mathcal{R}^{j_\mathbf{x}^k})$. 
The {\it spin network} basis elements are the following 
gauge invariant functions:
\begin{eqnarray} 
\psi_{\vec{\jmath},\mathbf{\vec{\pi}}}(U) 
   &=& \prod_{\mathbf{x}} 
       \Bigg[ \prod_{k=1}^{d}             
              \sum_{\alpha_\mathbf{x}^k,\beta_\mathbf{x}^k=1}^{
                   {\mathrm{dim}(j_\mathbf{x}^k)}} 
       ~\Rgen[j_\mathbf{x}^k ]{{\alpha}_\mathbf{x}^k}{
       {\beta}_\mathbf{x}^k}(U_k(\mathbf{x})) ~\Bigg] 
       \cdot
    \mathsf{I}_{\mathbf{x}}^{\Sg{\mathbf{\pi}_\mathbf{x}}}
    {}^\g{j_{\xm[d]}^1,\ldots,j_{\xm[d]}^d}_{
        \alpha_{\xm[1]}^1,\ldots,\alpha_{\xm[d]}^d}
    {}_\g{j_\mathbf{x}^1,\ldots,j_\mathbf{x}^d}^{
        \beta_\mathbf{x}^1,\ldots,\beta_\mathbf{x}^d}
~~ .
\label{eq:spinNET}
\end{eqnarray}
Using the unitarity of the representations $\Rgen[j]{\alpha}{\beta}(U)$ 
we have that the complex conjugated elements is given by:
\begin{eqnarray} 
\overline{\psi_{\vec{\jmath},\mathbf{\vec{\pi}}}(U)} 
   &=& \prod_{\mathbf{x}} 
       \Bigg[ \prod_{k=1}^{d}             
              \sum_{\alpha_\mathbf{x}^k,\beta_\mathbf{x}^k=1}^{
                   {\mathrm{dim}(j_\mathbf{x}^k)}} 
       ~\Rgen[j_\mathbf{x}^k ]{{\beta}_\mathbf{x}^k}{
       {\alpha}_\mathbf{x}^k}(U_k^{\gg{-1}}(\mathbf{x})) ~\Bigg] 
       \cdot
    \mathsf{I}^{\mathbf{x}}_{\Sg{\mathbf{\pi}_\mathbf{x}}}
    {}_\g{j_{\xm[d]}^1,\ldots,j_{\xm[d]}^d}^{
        \alpha_{\xm[1]}^1,\ldots,\alpha_{\xm[d]}^d}
    {}^\g{j_\mathbf{x}^1,\ldots,j_\mathbf{x}^d}_{
        \beta_\mathbf{x}^1,\ldots,\beta_\mathbf{x}^d}
~~ .
\label{eq:spinNETcc}
\end{eqnarray}
For the computation of the magnetic field term, we need the following 
integrals (with $j=1$ we denote the defining representation, i.e.,
$\Rgen[1 ]{{\alpha}}{{\beta}}(U)=U^\alpha_\beta$)
\begin{eqnarray}\label{p:INT2}
 \int dU~ \Rgen[j ]{\alpha}{\beta}(U)
          \Rgen[1 ]{\bar{\alpha}}{\bar{\beta}}(U)
          \Rgen[j']{\beta'}{\alpha'}(U^{-\!1})
&=& \sum_{\pi} \frac{
    \IUg[\pi]{}^\gg{j1}_{\beta \bar{\beta }}{}_\gg{j'}^{\beta'}
    \IDg[\pi]{}_\gg{j1}^{\alpha\bar{\alpha}}{}^\gg{j'}_{\alpha'}
  }{\IUg[\pi]{}^\gg{j1}_{\gamma\bar{\gamma}}{}_\gg{j'}^{\gamma'}
    \IDg[\pi]{}_\gg{j1}^{\gamma\bar{\gamma}}{}^\gg{j'}_{\gamma'}
   }
\\ \label{p:INT3}
 \int dU~ \Rgen[j ]{\alpha}{\beta}(U)
          \Rgen[1 ]{\bar{\beta}}{\bar{\alpha}}(U^{-\!1})
          \Rgen[j']{\beta'}{\alpha'}(U^{-\!1}) 
&=& \sum_{\pi} \frac{
    \IUg[\pi]{}^\gg{j}_{\beta }{}_\gg{1j'}^{\bar{\beta }\beta'}
    \IDg[\pi]{}_\gg{j}^{\alpha}{}^\gg{1j'}_{\bar{\alpha}\alpha'}
  }{\IUg[\pi]{}^\gg{j}_{\gamma}{}_\gg{1j'}^{\bar{\gamma}\gamma'}
    \IDg[\pi]{}_\gg{j}^{\gamma}{}^\gg{1j'}_{\bar{\gamma}\gamma'}
   } ~~.
\end{eqnarray}
The presence of the $\sum_\pi$ over a basis of possible three valent
intertwiners denotes the fact that for groups of ranks greater than
$1$ a given representation can appear more than once in the tensor 
product of two representations. This is not the case, as it is well
known, for the $SU(2)$ group. In this case no sum over $\pi$ appears.

Summarizing, our characterization of the Hilbert space of Lattice Gauge 
Theories require: 
\begin{enumerate}
\item   The determination and description of the set of all the 
        unitary inequivalent representations of the group $G$. By this,
        we mean their explicit form and the construction of a unique
        indexing $J[G]$ of them, i.e., the complete knowledge of the
        set  $\mathcal{RG} = \{  \mathcal{R}^{j} ~| j\in J[G] \}$.\item  The description of the space of all the intertwining
         operators $\mathcal{I}=\mathcal{I}(\mathcal{R}^{j_1}\otimes
         \ldots\otimes\mathcal{R}^{j_n},\mathcal{R}^{i_1}\otimes
         \ldots\otimes\mathcal{R}^{i_m})$, and the determination of 
         a basis $\mathsf{I}^\Sg{v}$ on it
         ($\mathsf{I}\in \mathcal{I} \leftrightarrow \mathsf{I}=\sum          c_{\Sg{v}} \mathsf{I}^\Sg{v}$).
          This involves the decomposition of an arbitrary representation
         in the tensor product of irreducible representations.
\end{enumerate}
In the follow we will see that in order to compute the matrix element
the only explicit function we will need are: (a) the value of the
quadratic Casimir invariant on the $\mathcal{R}^{j}$ representation:
$C_2[j]$; (b) the explicit values of the contraction of arbitrary
intertwiner matrix.  Without entering in details, extensively treated in 
literature \cite{Yutsis:1962,Citanovic:1984,Brink:1968,Vilenkin:1995},
we want to emphasize that the computation of the trace of intertwiner
matrices can be alway reduced to the computation of the sum of product
of Wigner's $6J$-symbols. Such elements, in particular, are very well
known for the $SU(2)$ group (see for example \cite{Yutsis:1962}), and
a quite extensive bibliography and collection of results exist for
other compact groups. See for example: \cite{deSwart:1963} for SU(3),
\cite{Shelepin:1965} for $SU(N)$ and \cite{Vilenkin:1995,Barut:1986} for
general overviews of known results.


\section{The matrix elements of the Hamiltonian operator}

We can perform the computation of the action of the 
Hamiltonian operator (\ref{def:HAM}) on the {\it spin-networks basis} 
(\ref{eq:SpinNetBasis}). In fact, the
basis vectors \ref{eq:spinNET} are eigenstate of the kinetic
term $H_E$, while the potential (magnetic) term is realized as 
a multiplicative operator, i.e.:  
\begin{eqnarray} \label{eq:HspinNET}
\langle \vec{\jmath}~',\mathbf{\vec{\pi}'}| 
\hat{H} | \vec{\jmath},\mathbf{\vec{\pi}}\rangle
&=&  \left( \frac{g^2}{2a} \sum_{\mathbf{x}}\sum_{k=1}^{d}
              C_2[j_\mathbf{x}^2]
       ~
     +\frac{1}{a g^2} N_P \right) 
     ~\langle \vec{\jmath}~',\mathbf{\vec{\pi}'}| 
           \vec{\jmath},\mathbf{\vec{\pi}}\rangle
\\
& & - \frac{1}{ag^2~2~{\rm dim}(U)}
\sum_{\mathbf{y}} \sum_{\substack{r,s=1\\r<s}}^d 
\bigg(
   \langle \vec{\jmath}~',\mathbf{\vec{\pi}'}| 
   U_{\mathbf{y},r,s} | \vec{\jmath},\mathbf{\vec{\pi}}\rangle
  +\langle \vec{\jmath},\mathbf{\vec{\pi}}| 
   U_{\mathbf{y},r,s} | \vec{\jmath}~',\mathbf{\vec{\pi}'}\rangle
\bigg)
\nonumber
\end{eqnarray}
where the only non diagonal terms is given by the expection 
values of the plaquette operator. 


\subsection{The plaquette operator} 

Using the integrals (\ref{p:INT2}-\ref{p:INT3}), it is straightforward
to compute the matrix elements of the plaquette operators in the spin
network basis. The final result, expressed as traces in the
intertwiner spaces associated to each vertex, is
\begin{eqnarray}
&& \langle \vec{\jmath}~',\mathbf{\vec{\pi}'}| 
U_{\mathbf{y},r,s} 
| \vec{\jmath},\mathbf{\vec{\pi}}\rangle
= \int \prod_{\mathbf{x},k} dU_k(\mathbf{x})
\overline{\psi_{\vec{\jmath}~',\mathbf{\vec{\pi}}'}(U)}
\psi_{\vec{\jmath},\mathbf{\vec{\pi}}}(U) 
U^{-\!1}_s{}^{\tau_1}_{\tau_4}(\mathbf{y}) 
U^{-\!1}_r{}^{\tau_4}_{\tau_3}(\mathbf{y}+a \mathbf{e}_s)
U_s{}^{\tau_3}_{\tau_2}(\mathbf{y}+a \mathbf{e}_r) 
U_r{}^{\tau_2}_{\tau_1}(\mathbf{y})
\nonumber \\ 
&& ~~~
 = \int 
   \prod_{\mathbf{x}} \prod_{k=1}^{d} \Bigg[ 
   dU_k(\mathbf{x})
   \sum_{\bar{\alpha}_\mathbf{x}^k,\bar{\beta}_\mathbf{x}^k=1}^{
                     {\mathrm{dim}(j_\mathbf{x}^k)}} 
   \sum_{\alpha_\mathbf{x}^k,\beta_\mathbf{x}^k=1}^{
                     {\mathrm{dim}(j_\mathbf{x}^{k\prime})}} 
   ~\Rgen[j_\mathbf{x}^k ]{\bar{\beta}_\mathbf{x}^k}{
                          {\bar\alpha}_\mathbf{x}^k}(U_k^{-\!1}(\mathbf{x})) 
   ~\Rgen[j_\mathbf{x}^{k\prime}]{{{\alpha}}_\mathbf{x}^k}{
                          {{\beta}}_\mathbf{x}^k}(U_k(\mathbf{x})) 
   ~\Bigg]
   \cdot
\\[2mm] 
&& ~~~~~
       \cdot 
       U^{-\!1}_s{}^{\tau_1}_{\tau_4}(\mathbf{y}) 
       U^{-\!1}_r{}^{\tau_4}_{\tau_3}(\mathbf{y}+a \mathbf{e}_s)
       U_s{}^{\tau_3}_{\tau_2}(\mathbf{y}+a \mathbf{e}_r) 
       U_r{}^{\tau_2}_{\tau_1}(\mathbf{y})
       \cdot 
    \mathsf{I}_{\mathbf{x}}^{\Sg{{\mathbf{\pi}_\mathbf{x}}}}
    {}^\g{j_{\xm[1]}^1,\ldots,j_{\xm[d]}^d
     }_{\bar{\alpha}_{\xm[d]}^1,\ldots,\bar{\alpha}_{\xm[d]}^d}
    {}_\g{j_\mathbf{x}^1,\ldots,j_\mathbf{x}^d
     }^{\bar{\beta}_\mathbf{x}^1,\ldots,\bar{\beta}_\mathbf{x}^d}
       \cdot
    \mathsf{I}^\mathbf{x}_\Sg{\mathbf{\pi}_\mathbf{x}'}
    {}_\g{j_{\xm[1]}^{1\prime},\ldots,j_{\xm[d]}^{d\prime}
     }^{\alpha_{\xm[1]}^1,\ldots,\alpha_{\xm[d]}^d}
    {}^\g{j_\mathbf{x}^{1\prime},\ldots,j_\mathbf{x}^{d\prime}
     }_{\beta_\mathbf{x}^1,\ldots,\beta_\mathbf{x}^d}
\nonumber 
\end{eqnarray}
Therefore, the result reduces to the computation of the trace of the
intertwing matrix. This can be done by using equations (\ref{p:INT1}),
(\ref{p:INT2}) and (\ref{p:INT3}). This result shows that the choice
of an explicit basis is irrelevant, as expected. All the indices of the 
intertwiner matrix elements are traced over
their complex conjugate, except the contractor in the lattice points
$\mathbf{y}$, $\yp[r]$, $\yp[s]$ and $\yp[r+s]$.  The corresponding
matrix elements are given by:
\begin{eqnarray}
&&
\langle \mathbf{\vec{\jmath}~'},\mathbf{\vec{\pi}'}| 
U_{\mathbf{y},r,s} | \mathbf{\vec{\jmath}},\mathbf{\vec{\pi}}\rangle
=  
  \prod_{\mathbf{x}} \prod_{k=1}^{d} \Bigg[ 
  \sum_{\bar{\alpha}_\mathbf{x}^k,\bar{\beta}_\mathbf{x}^k=1}^{
                         {\mathrm{dim}(j_\mathbf{x}^k)}} 
  \sum_{\alpha_\mathbf{x}^k,\beta_\mathbf{x}^k=1}^{
                         {\mathrm{dim}(j_\mathbf{x}^{k\prime})}} 
  \bigg]
  \Bigg[ \!\!
   \prod_{\substack{ (\mathbf{x},k) \neq  \\
           \neq (\mathbf{y},r),(\yp[r],s) \\
           ,(\yp[s],r),(\mathbf{y},s) }} \!\!
      \frac{\delta_{j_\mathrm{x}^k}^{j_\mathrm{x}^{k\prime}}
          }{\mathrm{dim}(j_\mathrm{x}^k)}
  \bigg]
\label{eq:completa} \\[3mm] &&
  \cdot \Bigg\{ \!\!
     \prod_{\substack{\mathbf{x}\neq \\
             \neq \mathbf{y},\yp[r]\\
             ,\yp[s],\yp[r+s]}} \!\! 
    \mathsf{I}_{\mathbf{x}}^{\Sg{\mathbf{\pi}_\mathbf{x}}}
    {}^\g{j_{\xm[1]}^{1},\ldots,j_{\xm[d]}^{d}}_{
        {\bar\alpha}_{\xm[1]}^1,\ldots,{\bar\alpha}_{\xm[d]}^d}
    {}_\g{j_\mathbf{x}^1,\ldots,j_\mathbf{x}^d}^{
        {\bar\beta}_\mathbf{x}^1,\ldots,{\bar\beta}_\mathbf{x}^d}
    ~\cdot~ 
    \mathsf{I}^\mathbf{x}_\Sg{\mathbf{\pi}_\mathbf{x}'}
    {}_\g{j_{\xm[d]}^{1\prime},\ldots,j_{\xm[d]}^{d\prime}}^{
        \alpha_{\xm[1]}^1,\ldots,\alpha_{\xm[d]}^d}
    {}^\g{j_\mathbf{x}^{1\prime},\ldots,j_\mathbf{x}^{d\prime}}_{
        \beta_\mathbf{x}^1,\ldots,\beta_\mathbf{x}^d}
   ~\cdot~ \left[ \prod_{k} 
           \delta_{\alpha_\mathbf{x}^k}^{\bar{\alpha}_\mathbf{x}^k}
   ~\cdot~ \prod_{k\neq r,s} 
           \delta_{\bar{\beta}_\mathbf{x}^k}^{{\beta}_\mathbf{x}^k}
   \right]
   \Bigg\}
\nonumber  \\[3mm] &&
  \cdot \sum_{\pi_1,\pi_2,\pi_3,\pi_4} \Bigg\{  \left[
     \frac{\IUg[\pi_1]{}^\gg{j_\mathbf{y}^r1
                  }_{\bar{\beta}_\mathbf{y}\delta_2}
                 {}_\gg{j_\mathbf{y}^{r\prime}
                 }^{\beta_\mathbf{y}}
      \IDg[\pi_1]{}_\gg{j_\mathbf{y}^r1}^{\bar{\alpha}_\mathbf{y}^r\delta_1}
                 {}^\gg{j_\mathbf{y}^{r\prime}
                  }_{{\alpha}_\mathbf{y}^r}
    }{\IUg[\pi_1]{}^\gg{j_\mathbf{y}^r1}_{\gamma_1\gamma_2}
                 {}_\gg{j_\mathbf{y}^{r\prime}}^{\gamma_3}
      \IDg[\pi_1]{}_\gg{j_\mathbf{y}^r1}^{\gamma_1\gamma_2}
                 {}^\gg{j_\mathbf{y}^{r\prime}}_{\gamma_3}
} \right]
\left[
\frac{\IUg[\pi_2]{}^\gg{j_\yp[r]^s1
                 }_{\bar{\beta}_\yp[r]^s\delta_3}
                 {}_\gg{j_\yp[r]^{s\prime}
                  }^{\beta_\mathbf{y}}
      \IDg[\pi_2]{}_\gg{j_\yp[r]^s1}^{\bar{\alpha}_\yp[r]^s\delta_2}
                 {}^\gg{j_\yp[r]^{s\prime}
                  }_{{\alpha}_\yp[r]^s}
    }{\IUg[\pi_2]{}^\gg{j_\yp[r]^r1}_{\gamma_1\gamma_2}
                 {}_\gg{j_\yp[r]^{r\prime}}^{\gamma_3} 
      \IDg[\pi_2]{}_\gg{j_\yp[r]^r1}^{\gamma_1\gamma_2}
                 {}^\gg{j_\yp[r]^{r\prime}}_{\gamma_3}
}\right]
\left[
\frac{\IUg[\pi_3]{}^\gg{j_\yp[r]^s}_{\bar{\beta}_\yp[r]^s }
                 {}_\gg{1j_\yp[r]^{s\prime}
                  }^{\delta_3\beta_\yp[r]^{s\prime}}
      \IDg[\pi_3]{}_\gg{j_\yp[r]^s}^{\alpha_\yp[r]^s}
                 {}^\gg{1j_\yp[r]^{s\prime}
                  }_{\delta_4\alpha_\yp[r]^{s\prime}}
    }{\IUg[\pi_3]{}^\gg{j_\yp[r]^s}_{\gamma_1}
                 {}_\gg{1j_\yp[r]^{s\prime}}^{\gamma_2\gamma_2}
      \IDg[\pi_3]{}_\gg{j_\yp[r]^s}^{\gamma_1}
                 {}^\gg{1j_\yp[r]^{s\prime}}_{\gamma_2\gamma_3}
} \right]
\nonumber \\[3mm] &&
\left[
\frac{\IUg[\pi_4]{}^\gg{j_\mathbf{y}^s}_{\bar{\beta}_\mathbf{y}^s }
                 {}_\gg{1j_\mathbf{y}^{s\prime}
                  }^{\delta_4\beta_\mathbf{y}^{s\prime}}
      \IDg[\pi_4]{}_\gg{j_\mathbf{y}^s}^{\alpha_\mathbf{y}^s}
                 {}^\gg{1j_\mathbf{y}^{s\prime}
                  }_{\delta_1\bar{\alpha}_\mathbf{y}^{s\prime}}
    }{\IUg[\pi_4]{}^\gg{j_\mathbf{y}^s}_{\gamma_1}
                 {}_\gg{1j_\mathbf{y}^{s\prime}}^{\gamma_2\gamma_3}
      \IDg[\pi_4]{}_\gg{j_\mathbf{y}^s}^{\gamma_1}
                 {}^\gg{1j_\mathbf{y}^{s\prime}}_{\gamma_2\gamma_3}
} \right]
    \mathsf{I}_{\mathbf{y}}^\Sg{\mathbf{\pi}_\mathbf{y}}
    {}^\g{j_{\xm[d]}^{1},\ldots,j_{\xm[d]}^{d}}_{
        \bar{\alpha}_{\xm[d]}^1,\ldots,\bar{\alpha}_{\xm[d]}^d}
    {}_\g{j_\mathbf{y}^1,\ldots,j_\mathbf{y}^d}^{
        \bar{\beta}_\mathbf{y}^1,\ldots,\bar{\beta}_\mathbf{y}^d}
    \cdot
    \mathsf{I}^\mathbf{y}_\Sg{\mathbf{\pi}_\mathbf{y}^\prime}
    {}_\g{j_{\ym[1]}^{1\prime},\ldots,j_{\ym[d]}^{d\prime}}^{
        \alpha_{\ym[1]}^1,\ldots,\alpha_{\ym[d]}^d}
    {}^\g{j_\mathbf{y}^{1\prime},\ldots,j_\mathbf{x}^{d\prime}}_{
        \beta_\mathbf{y}^1,\ldots,\beta_\mathbf{x}^d}
   ~\cdot~ \left[ \prod_{k} 
           \delta_{\alpha_\mathbf{x}^k}^{\bar{\alpha}_\mathbf{x}^k}
   ~\cdot~ \prod_{k\neq r,s} 
           \delta_{\bar{\beta}_\mathbf{x}^k}^{{\beta}_\mathbf{x}^k}
    \right] \cdot
\nonumber \\[3mm] &&
    \cdot
    \mathsf{I}_{\yp[r]}^{\Sg{\mathbf{\pi}_\yp[r]}}
    {}^\g{j_{\yp[r-1]}^{1},\ldots,j_{\yp[r-d]}^{d}}_{
         \bar{\alpha}_{\yp[r+1]}^1,\ldots,\bar{\alpha}_{\yp[r+d]}^d}
    {}_\g{j_\yp[r]^1,\ldots,j_\yp[r]^d}^{
        \bar{\beta}_\yp[r]^1,\ldots,\bar{\beta}_\yp[r]^d}
    \cdot
    \mathsf{I}^\yp[r]_\Sg{\mathbf{\pi}_\yp[r]^\prime}
    {}_\g{j_{\yp[r-1]}^{1\prime},\ldots,j_{\yp[r-d]}^{d\prime}}^{
        \alpha_{\yp[r+1]}^1,\ldots,\alpha_{\yp[r+d]}^d}
    {}^\g{j_\yp[r]^{1\prime},\ldots,j_\yp[r]^{d\prime}}_{
        \beta_\yp[r]^1,\ldots,\beta_\yp[r]^d}
   ~\cdot~ \left[ \prod_{k\neq r} 
           \delta_{{\alpha}_\yp[r-k]^k}^{\bar{\alpha}_\yp[r-k]^k}
   ~\cdot~ \prod_{k\neq s} 
           \delta_{\bar{\beta}_\yp[r]^k}^{{\beta}_\yp[r]^k}
   \right] \cdot
\nonumber \\[3mm] &&
    \cdot
    \mathsf{I}_{\yp[r+s]}^\Sg{\mathbf{\pi}_\yp[r+s]}
    {}^\g{j_{\yp[r-1]}^{1},\ldots,j_{\yp[r+s-d]}^{d}}_{
        \bar{\alpha}_{\yp[r+1]}^1,\ldots,\bar{\alpha}_{\yp[r+s-d]}^d}
    {}_\g{j_\yp[r+s]^1,\ldots,j_\yp[r+s]^d}^{
        \bar{\beta}_\yp[r+s]^1,\ldots,\bar{\beta}_\yp[r+s]^d}
    \cdot
    \mathsf{I}^\yp[r+s]_\Sg{\mathbf{\pi}_\yp[r+s]^\prime}
    {}_\g{j_{\yp[r+s-1]}^{1\prime},\ldots,j_{\yp[r+s-d]}^{d\prime}}^{
        \alpha_{\yp[r+s-1]}^1,\ldots,\alpha_{\yp[r+s-d]}^d}
    {}^\g{j_\yp[r+s]^{1\prime},\ldots,j_\yp[r+s]^{d\prime}}_{
        \beta_\yp[r+s]^1,\ldots,\beta_\yp[r+s]^d}
\nonumber \\[3mm] &&
   \cdot \left[ \prod_{k\neq r,s} 
           \delta_{{\alpha}_\yp[r+s-k]^k}^{\bar{\alpha}_\yp[r+s-k]^k}
   \cdot \prod_{k} 
           \delta_{\bar{\beta}_\yp[r+s-k]^k}^{{\beta}_\yp[r+s-k]^k}
   \right] \cdot
    \cdot
    \mathsf{I}_\yp[s]^\Sg{\mathbf{\pi}_\yp[s]}
    {}^\g{j_{\yp[s-1]}^{1},\ldots,j_{\yp[s-d]}^{d}}_{
        \bar{\alpha}_{\yp[s-1]}^1,\ldots,\bar{\alpha}_{\yp[s-d]}^d}
    {}_\g{j_\yp[s]^1,\ldots,j_\yp[s]^d}^{
        \bar{\beta}_\yp[s]^1,\ldots,\bar{\beta}_\yp[s]^d}
    \cdot
    \mathsf{I}^\yp[s]_\Sg{\mathbf{\pi}_\yp[s]^{\prime}}
    {}_\g{j_{\yp[s-1]}^{1\prime},\ldots,j_{\yp[s-d]}^{d\prime}}^{
        \alpha_{\yp[s-1]}^1,\ldots,\alpha_{\yp[s-d]}^d}
    {}^\g{j_\yp[s]^{1\prime},\ldots,j_\yp[r]^{d\prime}}_{
        \beta_\yp[s]^1,\ldots,\beta_\yp[s]^d}
\nonumber \\[3mm] &&
   ~\cdot~ \left[
           \prod_{k\neq r} 
           \delta_{{\alpha}_\yp[s-k]^k}^{\bar{\alpha}_\yp[s-k]^k}
   ~\cdot~ \prod_{k\neq r} 
           \delta_{\bar{\beta}_\yp[s]^k}^{{\beta}_\yp[s]^k}
   \right] 
\Bigg\}
\nonumber
\end{eqnarray}
where the quantities inside the square brackets $[...]$ come from the 
group integrations. Notice that once the intertwining matrices are 
specified, i.e., when the Clebsh-Gordan coefficients are explicitly given, 
the matrix elements are known. The only non trivial part in this 
computation is the choice of a convenient basis for the intertwining 
matrices. A natural choice is to use an orthonormal basis, i.e., 

\begin{equation}
    \sum_{\alpha_\mathbf{x}^k,\beta_\mathbf{x}^k=1}^{
                         {\mathrm{dim}(j_\mathbf{x}^{k})}} 
    \mathsf{I}_{\mathbf{x}}^{\Sg{\mathbf{\pi}_\mathbf{x}}}
    {}^\g{j_{\xm[1]}^{1},\ldots,j_{\xm[d]}^{d}}_{
        {\alpha}_{\xm[1]}^1,\ldots,{\alpha}_{\xm[d]}^d}
    {}_\g{j_\mathbf{x}^1,\ldots,j_\mathbf{x}^d}^{
        {\beta}_\mathbf{x}^1,\ldots,{\beta}_\mathbf{x}^d}
    ~\cdot~ 
    \mathsf{I}^\mathbf{x}_\Sg{\mathbf{\pi}_\mathbf{x}'}
    {}_\g{j_{\xm[d]}^{1},\ldots,j_{\xm[d]}^{d}}^{
        \alpha_{\xm[1]}^1,\ldots,\alpha_{\xm[d]}^d}
    {}^\g{j_\mathbf{x}^{1},\ldots,j_\mathbf{x}^{d}}_{
        \beta_\mathbf{x}^1,\ldots,\beta_\mathbf{x}^d}
    = \delta_\Sg{\mathbf{\pi}_\mathbf{x}'}^\Sg{\mathbf{\pi}_\mathbf{x}} ~~.
\end{equation}
In this way we have reduced the problem of the computation of
the matrix elements of the plaquette operator to the computation
of the trace of intertwining operators, i.e., of the trace of 
generalized Clebsh-Gordan coefficients. Now, it is well known 
that this is nothing more than the evaluation of specific 
Wigner's $nJ$-symbols and that the evaluation of a  
Wigner's $nJ$-symbols can be always reduced to the computation 
of a Wigner's $6J$-symbol. 

This means that what we really need to explicitly compute the matrix 
elements of the plaquette operator is just the knowledge of all the 
representations of the group $G$ and of the associated Wigner's 
$6J$-symbol.  In the case of the $SU(2)$ group these elements are 
known and standard references for this kind of 
computation are \cite{Yutsis:1962}
and \cite{Brink:1968}. A useful convention for doing
this is given by Penrose's binor calculus, 
where the distinction between the representations $\mathcal{R}^j$ and 
$\overline{\mathcal{R}^j}$
completely disappears \cite{DePietri:1997}.


\subsection{Matrix elements of the plaquette operator for 
         $SU(2)$ theory in 2+1 and 3+1 dimensions}

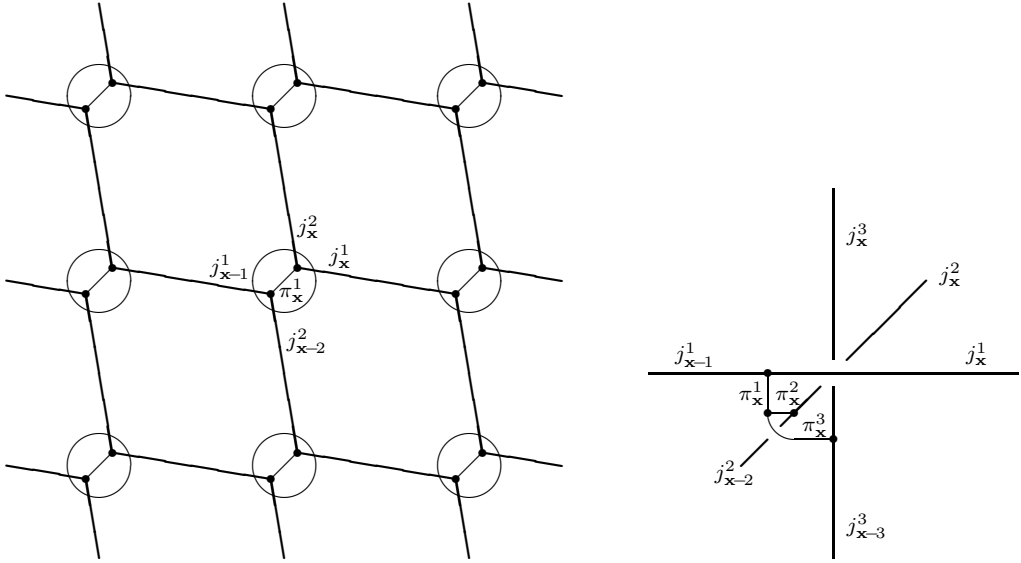
\begin{figure}
\def\BasicBlockTwoDim#1#2#3#4#5{{
\mbox{{\setlength{\unitlength}{1pt}
\begin{picture}(70,70)
   \put( 7,38){$\scriptstyle #1$}
   \put(36,10){$\scriptstyle #2$}
   \put(33,29){$\scriptstyle #3$}
   \put(52,42){$\scriptstyle #4$}
   \put(40,53){$\scriptstyle #5$} 
   \thinlines 
   \put(30,30){\circle*{3}}      \put(40,40){\circle*{3}}
   \put(35,35){\circle{25}}
   \put(30,30){\line( 1,1){10}}
   \thicklines 
   \put( 0,35){\line(6,-1){30}}   \put(70,35){\line(-6, 1){30}}   
   \put(35, 0){\line(-1,6){5}}    \put(35,70){\line( 1,-6){5}}   
\end{picture}}}  }}
\centerline{$
\mbox{\setlength{\unitlength}{1pt}
\begin{picture}(210,210)
   \put(  0,  0){\BasicBlockTwoDim{}{}{}{}{}}
   \put( 70,  0){\BasicBlockTwoDim{}{}{}{}{}}
   \put(140,  0){\BasicBlockTwoDim{}{}{}{}{}}
   \put(  0, 70){\BasicBlockTwoDim{}{}{}{}{}}
   \put( 70, 70){\BasicBlockTwoDim{j_{\xm[1]}^1}{j_{\xm[2]}^2}{
                    \pi^1_\mathbf{x}}{j_\mathbf{x}^1}{j_\mathbf{x}^2}}
   \put(140, 70){\BasicBlockTwoDim{}{}{}{}{}}
   \put(  0,140){\BasicBlockTwoDim{}{}{}{}{}}
   \put( 70,140){\BasicBlockTwoDim{}{}{}{}{}}
   \put(140,140){\BasicBlockTwoDim{}{}{}{}{}}
\end{picture}}
~~~~~ ~~~~~ 
\mbox{\setlength{\unitlength}{1pt}
\begin{picture}(140,140)
   \put( 10, 75){$\scriptstyle j_\xm[1]^1    $}
   \put(120, 75){$\scriptstyle j_\mathbf{x}^1$}
   \put( 25, 30){$\scriptstyle j_\xm[2]^2    $}
   \put(110,105){$\scriptstyle j_\mathbf{x}^2$}
   \put( 75, 10){$\scriptstyle j_\xm[3]^3    $}
   \put( 75,120){$\scriptstyle j_\mathbf{x}^3$}
   \put( 34, 60){$\scriptstyle \pi_\mathbf{x}^1$}
   \put( 48, 60){$\scriptstyle \pi_\mathbf{x}^2$}
   \put( 58, 48){$\scriptstyle \pi_\mathbf{x}^3$}
   \thinlines 
   \put(45,55){\circle*{3}}
   \put(45,55){\line(0,1){15}}    \put(45,70){\circle*{3}}
   \put(45,55){\line(1,0){10}}    \put(55,55){\circle*{3}}
   \put(70,55){\oval(50,20)[bl]}  \put(70,45){\circle*{3}}
   \thicklines 
   \put( 0,70){\line(1,0){140}}
   \put(35,35){\line(1,1){10}}     \put(50,50){\line(1,1){15}} 
   \put(105,105){\line(-1,-1){30}}
   \put(70, 0){\line(0,1){65}}     \put( 70,140){\line( 0,-1){65}}
\end{picture}}
$}
\caption{Graphical representation of (a possible) decomposition of a
  vertex in a two dimensional (left) and in a three dimensional
  (right) lattice and local description of two and three dimensional
  lattices. The solid dots denote the presence of a Wigner 3J
  symbol. The Wigner 3J symbol is the more symmetric normalized choice
  of the unique (in the case of the \protect{$SU(2)$} group) 3-valent
  intertwiner operator (invariant tensor).}
\label{fig:VERTEX}
\end{figure}

When the gauge group involved is $SU(2)$ the set of all unitary
irreducible representations (labelled by spins) and the space
of all the intertwining operators (generalized Clebsh Gordan) are well
known. In general, a basis on the space of intertwiners can be
specified by $2d-3$ additional {\it virtual} spins (see
\cite{Yutsis:1962} chapter II or \cite{DePietri:1997}).
That means that in 2+1 dimensions it is necessary 
to specify three spins 
$j_\mathbf{x}^1,j_\mathbf{x}^2,\pi^1_\mathbf{x}\in J[SU(2)]$ to
each lattice point $\mathbf{x}$ , while in 3+1 dimensions it is necessary 
to specify six spins $j_\mathbf{x}^1,j_\mathbf{x}^2,j_\mathbf{x}^3,
\pi_\mathbf{x}^{1},\pi_\mathbf{x}^2,\pi_\mathbf{x}^{3} \in J[SU(2)]$,
to each $\mathbf{x}$. Consequently, the spin network basis in
dimension $d=2$ is
\begin{eqnarray}
|\vec{\jmath};\vec{\pi}\rangle
&=& \prod_{\mathbf{x}} 
       \Rgen[j_\mathbf{x}^1]{{\alpha}_\mathbf{x}^1}{
       {\beta}_\mathbf{x}^1}(U_1(\mathbf{x})) 
       \Rgen[j_\mathbf{x}^2]{{\alpha}_\mathbf{x}^2}{
       {\beta}_\mathbf{x}^2}(U_2(\mathbf{x}))
  \cdot
   g^{\beta_\mathbf{x}^1\beta_\mathbf{x}^{1\prime}}_{(\!j_\mathbf{x}^1 \!)}
   g^{\beta_\mathbf{x}^2\beta_\mathbf{x}^{2\prime}}_{(\!j_\mathbf{x}^2 \!)}
  \cdot 
\\ & & ~~~~ \nonumber
  \cdot \ThreeJ{ 
        j_{\xm[1]}^1       }{ j_{\xm[2]}^2       }{ \pi^1_\mathbf{x}     }{  
        \alpha_{\xm[1]}^1  }{ \alpha_{\xm[2]}^2  }{ \gamma_\mathbf{x}  }
   g^{\gamma_\mathbf{x}\gamma_\mathbf{x}'}_{(\!\pi^1_\mathbf{x}\!)}
      \ThreeJ{ 
  \pi_\mathbf{x}    }{j_\mathbf{x}^1            }{j_\mathbf{x}^2            }{
  \gamma_\mathbf{x}'}{\beta_\mathbf{x}^{1\prime}}{\beta_\mathbf{x}^{2\prime}}
\label{eq:H2}
\end{eqnarray}
while in dimension $d=3$ is given by:
\begin{eqnarray}\label{eq:H3}
|\vec{\jmath};\vec{\pi}\rangle 
&=&\prod_{\mathbf{x}} 
       \Rgen[j_\mathbf{x}^1]{{\alpha}_\mathbf{x}^1}{
       {\beta}_\mathbf{x}^1}(U_1(\mathbf{x})) 
       \Rgen[j_\mathbf{x}^2]{{\alpha}_\mathbf{x}^2}{
       {\beta}_\mathbf{x}^2}(U_2(\mathbf{x}))
       \Rgen[j_\mathbf{x}^3]{{\alpha}_\mathbf{x}^3}{
       {\beta}_\mathbf{x}^3}(U_3(\mathbf{x}))
  \cdot
   g^{\beta_\mathbf{x}^1\beta_\mathbf{x}^{1\prime}}_{(\!j_\mathbf{x}^1 \!)}
   g^{\beta_\mathbf{x}^2\beta_\mathbf{x}^{2\prime}}_{(\!j_\mathbf{x}^2 \!)}
   g^{\beta_\mathbf{x}^3\beta_\mathbf{x}^{3\prime}}_{(\!j_\mathbf{x}^3 \!)}
\\ && ~~~~ \nonumber 
  \cdot 
  \ThreeJ{j_{\xm[1]}^1     }{j_{\mathbf{x}}^1            }{\pi_\mathbf{x}^1}{
          \alpha_{\xm[1]}^1}{\beta_{\mathbf{x}}^{1\prime}}{
          \gamma_\mathbf{x}^{1\prime}}
  \cdot
  \ThreeJ{j_{\xm[2]}^2     }{j_{\mathbf{x}}^2}{\pi_\mathbf{x}^2}{
          \alpha_{\xm[2]}^2}{\beta_{\mathbf{x}}^{2\prime}}{
          \gamma_\mathbf{x}^{2\prime}}
  \cdot
  \ThreeJ{j_{\xm[1]}^3     }{j_{\mathbf{x}}^3}{\pi_\mathbf{x}^3}{
          \alpha_{\xm[1]}^3}{\beta_{\mathbf{x}}^{3\prime}}{
          \gamma_\mathbf{x}^{3\prime}}
    \cdot
\\ & & ~~~~ \nonumber ~~ \cdot 
    g^{\gamma_\mathbf{x}^1\gamma_\mathbf{x}^{1\prime}
    }_{(\!\pi_\mathbf{x}^1\!)}
    g^{\gamma_\mathbf{x}^2\gamma_\mathbf{x}^{2\prime}
    }_{(\!\pi_\mathbf{x}^2\!)}
    g^{\gamma_\mathbf{x}^3\gamma_\mathbf{x}^{3\prime}
    }_{(\!\pi_\mathbf{x}^3\!)}
    \cdot
    \ThreeJ{\pi_\mathbf{x}^1     }{\pi_\mathbf{x}^2}{\pi_\mathbf{x}^3}{
            \gamma_{\mathbf{x}}^1}{\gamma_{\mathbf{x}}^2}{
            \gamma_{\mathbf{x}}^3}
\end{eqnarray}
where $\ThreeJ{j_1}{j_2}{j_3}{m_1}{m_2}{m_3}$ and 
$g_{(\!j\!)}^{m_1n_1}$ are the standard Wigner 3J symbol and the 
group metric on the irreducible representation $j$. A straightforward
direct computation shows that the norm of this states is given by:
\begin{equation}
\sqrt{\langle\vec{\jmath};\vec{\pi}| \vec{\jmath};\vec{\pi}\rangle}
= \sqrt{\prod_{\mathbf{x}} 
  \left(\prod_{k=1}^d \frac{1}{\mathrm{dim}(j_\mathbf{x}^k)}   \right)
 ~\left(\prod_{k=1}^{2d-3} 
  \frac{1}{\mathrm{dim}(\pi_\mathbf{x}^k)}\right)
  } 
\end{equation}

\begin{figure}[t]
\def\BlockTwoDim#1#2#3#4#5{{
\mbox{{\setlength{\unitlength}{1.1pt}
\begin{picture}(70,70)
   \put( 2,38){$\scriptstyle #1$}
   \put(38, 2){$\scriptstyle #2$}
   \put(33,29){$\scriptstyle #3$}
   \put(46,43){$\scriptstyle #4$}
   \put(40,58){$\scriptstyle #5$} 
   \thinlines 
   \put(30,30){\circle*{3}}      \put(40,40){\circle*{3}}
   \put(35,35){\circle{25}}
   \put(30,30){\line( 1,1){10}}
   \thicklines 
   \put( 0,35){\line(6,-1){30}}   \put(70,35){\line(-6, 1){30}}   
   \put(35, 0){\line(-1,6){5}}    \put(35,70){\line( 1,-6){5}}   
\end{picture}}}  }}
\centerline{$
\left\langle 
\begin{array}{c}\mbox{\setlength{\unitlength}{1.1pt}
\begin{picture}(140,140)
   \put(  0,  0){\BlockTwoDim{
                            }{
                            }{C_\mathbf{y}^1 = \pi_\mathbf{y}^{1\prime}    
                            }{Y_\mathbf{y}^1 = j_\mathbf{y}^{1\prime}
                            }{
                            }}
   \put( 70,  0){\BlockTwoDim{
                            }{C_\mathbf{y}^6 = j_\yp[1\!-\!2]^{2\prime}
                            }{Y_\mathbf{y}^6 = \pi_\yp[1]^{1\prime}         
                            }{C_\mathbf{y}^5 = j_\yp[1]^{1\prime}        
                            }{
                            }}
   \put(  0, 70){\BlockTwoDim{C_\mathbf{y}^2 = j_{\yp[2\!-\!1]}^{1\prime}
                            }{Y_\mathbf{y}^2 = j_\mathbf{y}^{2\prime}
                            }{Y_\mathbf{y}^3 = \pi_\yp[2]^{1\prime}
                            }{Y_\mathbf{y}^4 = j_\yp[2]^{1\prime}
                            }{C_\mathbf{y}^3 = j_\yp[2]^{2\prime}
                            }} 
   \put( 70, 70){\BlockTwoDim{
                            }{Y_\mathbf{y}^5 = j_{\yp[1]}^{2\prime}
                            }{C_\mathbf{y}^4 = \pi_\yp[1\!+\!2]^{1\prime}
                            }{
                            }{  
                            }}
\end{picture} }\end{array}
~~~\right|
~~
U_{\mathbf{y},1,2}
~~
\left|\begin{array}{c}\mbox{\setlength{\unitlength}{1.1pt}
\begin{picture}(150,140)
   \put(  0,  0){\BlockTwoDim{
                            }{
                            }{C_\mathbf{y}^1 = \pi_\mathbf{y}^1 
                            }{X_\mathbf{y}^1 = j_\mathbf{y}^1    
                            }{
                            }}
   \put( 70,  0){\BlockTwoDim{
                            }{C_\mathbf{y}^6 = j_\yp[1\!-\!2]^2   
                            }{X_\mathbf{y}^6 = \pi_\yp[1]^1   
                            }{C_\mathbf{y}^5 = j_\yp[1]^1        
                            }{
                            }}
   \put(  0, 70){\BlockTwoDim{C_\mathbf{y}^2 = j_{\yp[2\!-\!1]}^1 
                            }{X_\mathbf{y}^2 = j_\mathbf{y}^2    
                            }{X_\mathbf{y}^3 = \pi_\yp[2]^1       
                            }{X_\mathbf{y}^4 = j_\yp[2]^1         
                            }{C_\mathbf{y}^3 = j_\yp[2]^2          
                            }} 
   \put( 70, 70){\BlockTwoDim{
                            }{X_\mathbf{y}^5 = j_{\yp[1]}^2       
                            }{C_\mathbf{y}^4 = \pi_\yp[1\!+\!2]^1   
                            }{
                            }{  
                            }}
\end{picture}}\end{array}
~~~\right\rangle
$}
\caption{Graphical representations of the computation of the 
   matrix elements of the plaquette operator 
   \protect{$\langle\mathbf{\vec{\jmath}~'},\mathbf{\vec{\pi}'}| 
   U_{\mathbf{y},1,2}|\mathbf{\vec{\jmath}},\mathbf{\vec{\pi}}\rangle$}.
   }\label{fig:2DIM}
\end{figure}
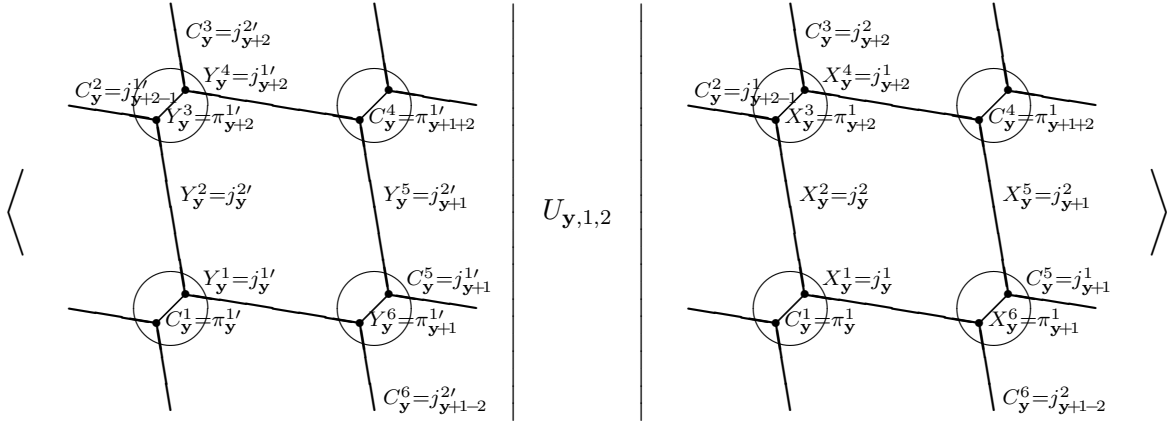

Using the explicit values of the $SU(2)$ Clebsh-Gordan coefficients,
the matrix elements of the Hamiltonian in equation (\ref{eq:completa})
can be computed. The calculation and the resulting expressions in two
and three dimensions are almost identical. The three dimensional case
has simply more indices to sum upon. The prototype of the generic
computation is indeed the two dimensional one. The expression
(\ref{eq:completa}) for the matrix elements $\langle
\mathbf{\vec{\jmath}}~',\mathbf{\vec{\pi}}'| U_{\mathbf{y},1,2} |
\mathbf{\vec{\jmath}},\mathbf{\vec{\pi}}\rangle$ becomes (in the basis
of equation (\ref{eq:H2}) and using the notation of figure
\ref{fig:2DIM}):
\begin{eqnarray}
\langle \mathbf{\vec{\jmath}~'},\mathbf{\vec{\pi}'}| 
U_{\mathbf{y},1,2} | \mathbf{\vec{\jmath}},\mathbf{\vec{\pi}}\rangle
&=&  
  \Bigg[ \!\!
   \prod_{\substack{ (\mathbf{x},k) \neq  \\
           \neq (\mathbf{y},1),(\yp[1],2) \\
           ,(\yp[2],1),(\mathbf{y},2) }} \!\!
      \frac{\delta_{j_\mathrm{x}^k}^{j_\mathrm{x}^{k\prime}}
          }{\mathrm{dim}(j_\mathrm{x}^k)}
  \bigg]
  \cdot \Bigg\{ \!\!
     \prod_{\mathbf{x}\neq \yp[1],\yp[2]} \!\! 
      \frac{\delta_{\pi_\mathrm{x}^1}^{\pi_\mathrm{x}^{1\prime}}
          }{\mathrm{dim}(\pi_\mathrm{x})}
   \Bigg\}
\label{eq:completa2} \\
& &  
 ~~~
 \prod_{i=1}^n 
   ~ 
   (-1)^\frac{ X^i_\mathbf{y}+Y^i_\mathbf{y} +
             + Y^{i\!+\!1}_\mathbf{y}
             + X^{i\!+\!1}_\mathbf{y}
             +C^i_\mathbf{y} +1 }{2}
   ~
   \SixJ{X^i_\mathbf{y}}{Y^i_\mathbf{y}}{1
       }{Y^{i\!+\!1}_\mathbf{y}}{X^{i\!+\!1}_\mathbf{y}}{C^i_\mathbf{y}}
\nonumber
\end{eqnarray} 
where
\begin{eqnarray}
\SixJ{j_1}{j_2}{j_3}{j_4}{j_5}{j_6} 
  &=& 
    \prod_{k=1}^6 \left[\sum_{m_k,n_k=-j_k}^{j_k}\right]
    g_\g{j_1}^{m_1n_1} g_\g{j_2}^{m_2n_2} g_\g{j_3}^{m_3n_3}
    g_\g{j_4}^{m_4n_4} g_\g{j_5}^{m_5n_5} g_\g{j_6}^{m_6n_6}  
\\[2mm] & & ~~
      \ThreeJ{j_1}{j_2}{j_3}{m_1}{m_2}{m_3}
      \ThreeJ{j_1}{j_5}{j_6}{n_1}{m_5}{m_6}
      \ThreeJ{j_4}{j_5}{j_3}{m_4}{n_5}{n_3}
      \ThreeJ{j_4}{j_2}{l_6}{n_4}{n_3}{n_6}
\nonumber 
\end{eqnarray}
is the Wigner $6J$-symbol.
>From the previous expression follows that the only elements different
from zero are those with equal $\mathbf{\vec{\jmath}}$ and
$\mathbf{\vec{\pi}}$ and unequal $\vec{X}_\mathbf{y}$ and
$\vec{Y}_\mathbf{y}$ (see figure \ref{fig:2DIM}). Indeed they must
differ exactly by 
$\epsilon_i={X}^i_\mathbf{y}-{Y}^i_\mathbf{y}=\pm \frac{1}{2}$ for
$i=1,...,n$ ($n=6$ for $d=2$ and $n=12$ for $d=3$). The explicit
values of the matrix elements are
\begin{equation} \label{eq:PLAQUETTESU2}
\frac{\langle \mathbf{\vec{\jmath}~'},\mathbf{\vec{\pi}'}| 
      U_{\mathbf{y},1,2} 
      | \mathbf{\vec{\jmath}},\mathbf{\vec{\pi}}\rangle
}{\sqrt{
  \langle\vec{\jmath}~';\vec{\pi}'|\vec{\jmath}~';\vec{\pi}'\rangle
  \langle\vec{\jmath};\vec{\pi}|\vec{\jmath};\vec{\pi}\rangle
}}
= \frac{
    (-1)^{\sum_{i=1}^n \left( \left|\epsilon_i-\epsilon_{i\!+\!1}\right|
                     + \frac{ C^i_\mathbf{y} }{2} \right)}
  }{
    \sqrt{\prod_{i=1}^n 
          \left( 2 X^i_\mathbf{y}+ 1 \right) 
          \left( 2 Y^i_\mathbf{y}+ 1 \right) }
  }
 \prod_{i=1}^n ~ 
   R\left[{ \begin{array}{cc} 
            X^i_\mathbf{y} & X^{i\!+\!1}_\mathbf{y} \\  
            Y^i_\mathbf{y} & Y^{i\!+\!1}_\mathbf{y}   
    \end{array},C^i_\mathbf{y} }\right] 
\end{equation}
where
\begin{equation}
   R\left[{ \begin{array}{cc} 
            X^i_\mathbf{y} & X^{i\!+\!1}_\mathbf{y} \\  
            Y^i_\mathbf{y} & Y^{i\!+\!1}_\mathbf{y}   
    \end{array},C^i_\mathbf{y} }\right] 
  = \left\{ \begin{array}{lcl} 
    \sqrt{\frac{1-2 C^i_\mathbf{y}
               +X^i_\mathbf{y}+X^{i\!+\!1}_\mathbf{y} 
               +Y^i_\mathbf{y}+Y^{i\!+\!1}_\mathbf{y} 
          }{2} ~
          \frac{3+2 C^i_\mathbf{y}
               +X^i_\mathbf{y}+X^{i\!+\!1}_\mathbf{y} 
               +Y^i_\mathbf{y}+Y^{i\!+\!1}_\mathbf{y} 
          }{2} 
    }
    & &\text{if~} \left|\epsilon_i-\epsilon_{i\!+\!1}\right|=0
    \\[3mm] 
    \sqrt{\frac{1+2 C^i_\mathbf{y}
               +X^i_\mathbf{y}-X^{i\!+\!1}_\mathbf{y} 
               +Y^i_\mathbf{y}-Y^{i\!+\!1}_\mathbf{y} 
          }{2} ~
          \frac{1+2 C^i_\mathbf{y}
               -X^i_\mathbf{y}+X^{i\!+\!1}_\mathbf{y} 
               -Y^i_\mathbf{y}+Y^{i\!+\!1}_\mathbf{y} 
          }{2} 
    }
    & &\text{if~} \left|\epsilon_i-\epsilon_{i\!+\!1}\right|=1
    \end{array}\right. 
\end{equation}


\section{Numerical vs. Analytic solutions}

In the previous sections we have shown how to map Lattice Gauge
Theories (based on arbitrary compact group) onto a well defined
``classical'' problem of quantum mechanics.  The problem of the
determination of the particle content of the theory and of the
$\beta$-function (imposing scaling conditions on the energy gaps) is
thus solved once the eigenvalues-eigenvectors of the theory are
known. Of course, this doesn't mean that we can straightforwardly give
such solution.

The main difficulties are two. First, the imposition of the gauge
invariance conditions, i.e., the labeling of the links with
inequivalent irreducible representations and of the vertices with a
basis of intertwining operators, give constraints which become harder
to deal with increasing volume. Since one has to simultaneously
satisfy all of them, and their interplay depends on the boundary
conditions, the determination of the explicit basis in infinite
(physical) volume is, although possible, not an easy task.  Second,
the associated spectral problem is not of obvious solution, and there
is no general algorithm which can diagonalize a given hermitian
operator in a reasonable time.

One can then follow two main guidelines. The first is to exploit
``brute-force'' numerical solutions in finite volume $V$ (IR-cutoff),
considering a finite dimensional subspace of the full Hilbert space,
e.g., restricting the analysis to a maximum allowed spin $\Lambda$ on
each lattice site (UV-cutoff). This procedure has the drawback not to
respect group symmetries. A similar, group invariant, cutoff can be
implemented considering a $q$-deformed gauge group
\cite{Burgio:1999}. In fact, when q is chosen such that $q^{n}=1$, the
number of irreducible representations is finite ($n=\Lambda +2$)
\cite{Vilenkin:1993c}. The main problem with such approaches is that
the dimension of the Hilbert space grows very rapidly with $\Lambda$
and $V$.  A rough estimation gives, in $d$ dimensions,
$dim(\Lambda,V)\simeq k~ \Lambda^{(2d-3)V}$, thus making
extrapolations to physical volume unlikely to be obtained.  The second
approach is to study the general properties of the problem at hand,
implementing symmetries and determining other conserved quantities
(i.e. observables commuting with the Hamiltonian).  This in order to
give at least constraints on the spectrum and on the form of the
eigenstates, thus reducing the set of states to be considered in the
diagonalization procedure.  Of course, a combination of analytical and
numerical techniques seems the most sensible thing to do.

As a final remark, let us show, for the sake of consistency, that we
correctly recover the known solutions for the vacuum state in the two
limits $g\rightarrow\infty$ and $g\rightarrow 0$.  In fact, can be 
straightforwardly checked that the vectors
\begin{equation}
  |0;g,a,L \rangle = \sum_{\vec{\jmath},\vec{\pi}} 
            c^\gg{0;g,a,L}_{\vec{\jmath},\vec{\pi}}
         \frac{|\vec{\jmath};\vec{\pi}\rangle}{
         \sqrt{\langle\vec{\jmath};\vec{\pi}|
               \vec{\jmath};\vec{\pi}\rangle}}~~, 
\end{equation}
where 
\begin{eqnarray}
c^\gg{0;+\infty,a,L}_{\vec{\jmath},\vec{\pi}}
&=&  \prod_{\mathbf{x}} \prod_{k} 
     \delta^{j_\mathbf{x}^k}_0 \delta^{\pi_\mathbf{x}^k}_0
\\
c^\gg{0;0,a,L}_{\vec{\jmath},\vec{\pi}}
&=&  \prod_{\mathbf{x}} \prod_{k} 
     \sqrt{\mathrm{dim}(j_\mathbf{x}^k)~\mathrm{dim}(\pi_\mathbf{x}^k)}
\end{eqnarray}
are eigenvectors of eigenvalues zero of the positive defined Hamiltonian operator
i.e., are proportional to the vacuum. Notice that the first result is trivial. 
The second, on the other hand, is  a consequence of the character expansion of the 
Dirac-$\delta$ function on the group and of the Bidenharm-Eliot identity.


\section{Outlook and conclusion}

In this work we showed that Peter-Weyl theorem gives a complete
characterization of the physical (gauge invariant) Hilbert space of
pure lattice gauge theories for any compact gauge group.  The
characterization is made in term of the the full set of irreducible
representations of the group and of a complete basis of the space of
intertwining operators.  Notice that such formalism can be
straightforwardly generalized to full gauge theories, where one can
associate a particle transforming according to the local gauge group at
each vertex \cite{Burgio:1999}.

In particular, we showed that, once a particular basis on the spaces
of intertwiner operator is selected, the matrix elements of the
Hamiltonian operator (\ref{eq:HspinNET},\ref{eq:completa}) are well
defined intertwiner contractions, i.e, can be expressed in terms of
the the evaluation of Wigner's $nJ$-symbol of the group $G$.  Moreover,
in the case of $SU(2)$ gauge group in 2+1 and 3+1 dimension we
derived a complete algebraic expression for such matrix elements
(\ref{eq:PLAQUETTESU2}).  These are our main results.

Recent results in group theory \cite{Vilenkin:1995,Archer:1992} allow us to 
think that the case of $SU(3)$ gauge group is not hopeless.

Extensions of such formalism to other theories, such as 2-dimensional
$\sigma$-models and $CP^N$ models are under investigations 
\cite{BurgioEtAl:1999}.

\vspace{.5cm}
\noindent {\sc Acknowledgment:}\\
We want to thank E. Onofri and F. Di Renzo for helpful 
and enlightening discussions. 
The work of R.D.P. at CPT Marseille is supported by a Dalla Riccia
Fellowship.  The work of H.A.M., L.F.U. and J.D.V. has been partially
supported by the CONACyT grant No. 3141P-E9608 and
grant DGAPA-UNAM-IN100397.

\vspace{1.0cm}


\end{document}